\documentclass[12pt,preprint]{aastex}

\shorttitle{Activity phases and Intermediate-degree mode frequencies}
\shortauthors{Jain, Tripathy and Hill}

\begin{document}

\title{Solar Activity Phases and Intermediate-degree Mode Frequencies}

\author{Kiran Jain, S. C. Tripathy and F. Hill}
\affil{Global Oscillation Network Group, National Solar Observatory, 
950 N Cherry Avenue, Tucson, AZ 85719, USA}
\email{kjain@noao.edu, stripathy@noao.edu, fhill@noao.edu}

\begin{abstract}
We analyze intermediate degree {\it p}-mode eigenfrequencies measured 
by GONG and MDI/SOHO over a solar cycle to study the source of their 
variability. We carry out a correlation analysis
between the change in frequencies and several measures of the Sun's magnetic 
activity that are sensitive to  changes at different levels in the solar atmosphere.
The observations span a period of about 12 years starting from mid-1996 
(the minimum of cycle 23) to  early-2008 (near minimum of cycle 24), 
corresponding to a nearly complete solar activity cycle. We demonstrate 
that the frequencies do vary in phase with the solar activity indices, however 
the degree of correlation differs from phase to phase of the
cycle. During 
the rising and declining phases, the mode frequency shifts are strongly 
correlated with the activity proxies whereas during the high-activity 
period, the shifts have significantly lower correlation with all 
activity proxies, except for the 10.7-cm radio flux. In particular, 
the proxies that are only influenced by the variation of the strong component
 of the magnetic field in the photosphere have a much lower correlation at the 
 high-activity period.  On the other hand, the shifts are  better 
 correlated with the proxies sensitive to  changes in the weak component
  of the magnetic field. Our correlation analysis suggests that 
more than 90\% of the variation  in the  oscillation frequencies in all
activity phases 
can be explained by changes in both components of the magnetic field. 
 Further, the slopes obtained from the linear regression analysis also 
 differ from phase to phase and show a strong correlation with the correlation 
 coefficients between frequency shifts and solar activity.
  
\end{abstract}


\section{Introduction}

Understanding the source of the varying solar oscillation eigenfrequencies has
been a topic of interest since the mid-eighties. A decrease in low-degree mode 
frequencies between 1980 (near solar maximum) and 1984 (near solar minimum) was 
initially reported by \citet{woo85}. Later, using Big Bear Solar Observatory data, 
\citet{woo91} showed that the frequencies were strongly correlated with the magnetic 
field strength. The variation  of the frequencies with the changing magnetic 
activity is now well established. Such changes affect both the central frequencies 
and the frequency splittings of low-degree  \citep{david04, wjc07} as well as 
intermediate-degree modes \citep{bachmann93, jain03, 
antia03, dzi05} and exhibit a good correlation between the mode frequencies 
and the solar activity. The high-degree mode frequencies calculated using 
the Dynamics Program data of the Michelson Doppler Imager (MDI) on board
{\it Solar and Heliospheric Observatory (SOHO)} also
 clearly show a solar cycle variation \citep{rs06}.

Detailed studies have shown that the mode frequencies do not always
follow the same paths during the rising and the declining phases of the solar activity 
\citep{jr98, sct01}. In addition, using the continuous oscillation data from
the Global Oscillation Network Group (GONG) for the rising phase of cycle 23, 
  \citet{howe99} demonstrated that, although the average frequency shift is
correlated well with measures of surface magnetic activity integrated over the 
whole disk, the even-order splitting coefficients had a higher correlation 
with the corresponding coefficients of a Legendre polynomial decomposition of 
the surface magnetic field than with the total flux. A recent analysis of the 
year-wise distribution of the frequency shifts with the change in activity indices 
showed that both the linear-regression slopes and the magnitude of the correlation 
varied from year to year \citep{sct07}. Their analysis further demonstrated that 
there were significant differences between long-term and short-term variations 
and concluded that averages of short-term variations do not accurately reflect 
the long-term variations and vice versa. The analysis of \citet{wjc07} for low 
degree modes involving three solar cycles also indicate that the frequency shifts 
are primarily sensitive to the weak component of the magnetic flux.   

Thus, the relationship between oscillation frequencies and solar activity
indicators seems to be more complicated than the linear relation
 generally demonstrated in earlier studies \citep{woo91, bachmann93, 
jain03}. With access to the high quality uniform data from GONG and MDI 
 instruments for a nearly complete solar cycle, it is timely to revisit the
temporal variability of the frequencies with the progression of solar 
activity cycle to understand the detailed relationship between them. 
In this paper, we closely examine the relationship between the frequency shifts 
of the intermediate degree modes and nine activity proxies for cycle 
23. We further divide the span of the activity cycle into three phases and  
compare the results obtained for different phases with the entire activity
cycle. The frequency and activity data used here are presented in Section 2. We
present results for individual data sets and their comparison in Section 3.
 A possible scenario to explain the variation in oscillation frequencies
is discussed in Section 4 and the main findings are summarized in Section 5. 

\section{Data}

\subsection{Mode Frequencies}

The analysis presented here uses intermediate-degree mode data
sets obtained from the GONG\footnote{ftp://gong.nso.edu/data/} and
 MDI\footnote{http://quake.stanford.edu/~schou/anavw72z/}, covering a period of
about 12 years i.e. a nearly complete solar cycle  starting from 1996 May 1.  
 Each  data set consists of centroid frequencies $\nu_{n,\ell}$ and splitting
coefficients $a_i$ and each ${n,\ell}$ multiplet is represented by a polynomial expansion
\begin{equation}
\nu_{n,\ell,m} = \nu_{n,\ell} + \sum_{i=1}^{i_{max}} a_i(n,\ell)
P_i^{(\ell)}(m),
\end{equation}
where $P_i^{(\ell)}(m)$'s are orthogonal polynomials of degree $i$  and
${i_{max}}$ is the number of $a$ coefficients used in determining 
frequencies \citep{schou92}.The remaining symbols in equation (1) have 
their usual meanings. The fitting is done using ${i_{max}}$ = 9 for GONG and 18
for MDI.  
 
 For the study reported here, we cover a period from  the minimum of
solar cycle 23 (1996 May) to the near minimum period of solar cycle 24 (2008
February). The continuous GONG data sets are centered at GONG Month (GM)
 12 to 124 while MDI data have two gaps in 1998--99 due to the breakdown of the
satellite. It should be also noted that each 108-day GONG data set 
starts 36 days later than the previous one having an overlap of 72 days while
the  72-day MDI data sets are non-overlapping.  
The MDI data have been obtained from the medium-$\ell$ program that
provides almost continuous coverage to modes with $\ell \le$ 300. 
Various other details of the data sets are given in Table~I and duty cycles
are plotted in Figure~1. We further note that the GONG data sets have relatively
 low duty cycle as compared to the MDI. The number of modes present in each GONG 
data set varies between 1700--1800, but only one-third of these modes are 
observed in all 113 overlapping data sets. Similarly, each MDI data set has 
approximately 1900 modes, which reduces to 885 common modes in all 58 data sets.  
The common mode sets in GONG and MDI are shown in Figure 2(a-b).   It should be 
noted that all these common modes are not successfully fitted for both projects, 
this might be due to the consequence of finite lifetime and stochastic nature of 
the modes. The common modes observed in both data sets are shown in Figure 2c.

\subsection{Proxies for Solar Activity}

To study the changes in mode frequencies with solar activity, 
we have used nine proxies of solar activity probing the solar atmosphere
at different heights. Each activity index is briefly summarized below; 
\begin{enumerate}
\item { $R_{\rm I}$, the international sunspot number computed from the 
total number of spots visible on the face of the sun ($N$) and the number of
groups ($G$) into which they cluster. $R_{\rm I}$ is equal to $10G + N$. }

\item {KPMI, the line of sight magnetic field strength  observed with 1 arc-sec
pixels using the photospheric \ion{Fe}{1} line at the Kitt Peak Vacuum
Telescope  and averaged over the full disk \citep{livingston76}.}

\item {MWSI, the Mt. Wilson sunspot index calculated using Mt. Wilson
magnetograms taken at the 150-Foot Solar Tower.  MWSI is determined by summing    
 the absolute values of all pixels with a magnetic field strength ($>$ 100 G)
which is then divided by the total number of pixels (regardless of magnetic 
field strength) in the magnetogram \citep{ulrich91}.}
 
\item { $F_{10} $, the integrated emission from the solar disc at 2.8 GHz i.e.
 10.7 cm wavelength. It mainly represents the contributions from sunspots and 
radio plages in the upper chromosphere in addition to the 
quiet-Sun background emission \citep{covington69,kundu65}.}

\item {\ion{He}{1}, the equivalent width of the \ion{He}{1} 1083 nm solar 
absorption line. It is affected by photo-ionization radiation from the 
upper transition region and corona. The index is calculated by measuring 
 the change in temperature in transition region \citep{harvey84}. } 
 
\item {MPSI, the magnetic plage strength index calculated in the same manner as
the MWSI but the absolute value of the magnetic field strength is summed for 
fields between 10~G and 100~G. }

\item {\ion {Mg}{2},
 the \ion{Mg}{2} core-to-wing ratio  derived by taking the ratio of the h and
 k lines of the solar \ion{Mg}{2} feature at 280 nm to the background or wings
at approximately 278 nm and 282 nm.  The h and k lines are variable chromospheric
emissions while the background emissions are more stable. This ratio is a 
robust measure of chromospheric activity, mainly for solar UV and EUV 
emissions \citep{ Viereck01}.}

\item {TSI,  the total solar irradiance, describes the radiant energy emitted
by the Sun over all wavelengths that falls each second on 1 square meter 
outside the earth's atmosphere. It is affected by the changes in both 
photospheric and  chromospheric magnetic structures. We use observations
made by VIRGO instrument on board {\it SOHO} \citep{cf97}.  }
 
\item {CI, the coronal index 
measures the total energy emitted by the
sun's outermost atmospheric layer (the corona) at a wavelength of 530.3 nm  
\citep{Rybansky94}}.

\end{enumerate}

It is worth mentioning that some activity indices are direct measurements
of weak, strong or both components of the magnetic field at different heights 
in the solar atmosphere while others are surrogates of magnetic measurements.
The strongest fields are concentrated in flux tubes in sunspots while the larger area
around sunspots harbors active regions and their plages contain small-scale
pores with relatively weak field strength.  The weak components are
more widely distributed in latitude than the strong components
 which are confined to the activity belts.  Since flux tubes expand like a
canopy in chromosphere above the photospheric footpoints, the field strength
decreases with increasing height. 

Table 2 summarizes each activity index, including the  spectral line of
observation, the period of observation  used in the analysis, the dominant component of
 the magnetic field, and the source of the data. A mean activity, $I$, is 
computed for each activity index over the same temporal intervals corresponding 
to the individual frequency measurements.   This mean value is used in 
subsequent calculations. 
\section{Analysis and Results} 

\subsection{Temporal variation in frequencies}

The centroid frequency shift $\delta \nu$ is calculated from the relation
\begin{equation}
\delta\nu(t)  = 
{\sum_{n,\ell}\frac{Q_{n,\ell}}{\sigma_{n,\ell}^2}\delta\nu_{n,\ell}(t)}/{\sum_{n,\ell}
\frac{Q_{n,\ell}}{\sigma_{n,\ell}^2}} ,
\end{equation}
where $\delta\nu_{n,\ell}(t)$ is the change in the measured frequency
with respect to a reference frequency, $\sigma_{n,\ell}$ is the uncertainty 
in the frequency measurement for a given $n$, $\ell$ multiplet 
 and  $Q_{n,\ell}$ is the inertia ratio as defined by \citet{jcd91}. 
 There are several definitions for selecting a reference frequency, which
have been discussed in detail by \citet{howe02}. For a given mode, 
we adopt an approach  of taking a temporal mean  over all 
the available data sets. It has  also been shown earlier that the frequency 
shift is strongly dependent on the frequency and increases with increasing  
 frequency (e.g. see Figure 3 of \citet{jain00}). We therefore restrict 
our analysis to common modes observed in all data sets to obtain an unbiased variation. 

 We use 235 modes observed in all GONG and MDI data sets in the degree range 
 of 21 $\le \ell \le$ 147, radial order range of 1 $\le n \le$ 16 and the
frequency range of 1.5 $\le \nu \le$ 4.0 mHz. These modes are shown in Figure~2c. 
The reference frequencies are calculated by taking an average of the frequencies 
of a particular multiplet ($n$,$\ell$)  from all of the available  GONG 
and MDI data sets. Figure~3 shows the temporal variation of frequency shift for
 both data sets. The dashed line  depicts the variation of the solar activity  
represented by radio flux, $F_{10}$. It is clearly seen that the frequencies vary 
in phase with the solar activity as shown in earlier studies (e.g.
\citet{jain03, wjc07}).   We also note that there is a systematic offset 
between GONG and MDI frequency shifts, which is consistent with the findings 
of \citet{howe02}. The MDI frequency shifts are consistently higher than 
the GONG and this deviation is relatively large at the low activity periods. 
However, the overall shifts from the minimum to the maximum in both cases 
are approximately same.  The  discrepancy observed might be due to the 
fact that the GONG and the MDI pipelines consider time series of 
different lengths and use different mathematical methods to determine 
the mode frequencies. A detailed correlation analysis between the frequency 
shifts and solar activity is presented in the next section.

\subsection{Correlation with solar activity} 
\subsubsection{Solar cycle variation}

The temporal variations in  frequency shifts with various measures of the 
solar activity are plotted in Figures~4  and 5 for GONG and MDI data sets, 
respectively. Although the frequency shifts follow the general trend of the solar 
activity, there are variations between different activity indices. For
example, we  note discrepancies between $\delta\nu$, and $R_{\rm I}$, MWSI and
TSI at  high activity period. We examine this variation by calculating the 
correlation between them. This is displayed in Table~3, where linear ($r_P$) and
rank ($r_S$) correlation coefficients between frequency shifts and the
measures of solar activity are tabulated. The linear correlation coefficient is
an estimate of the linear relationship between the two sets and the rank 
correlation shows whether  one set of numbers has any effect on another set of 
numbers. The high values of $r_P$ ($\ge 0.9$) clearly indicate a
linear relationship between two quantities. Therefore, we fit the shifts
against activity using linear expression 
 \begin{equation}
 \delta\nu = a + bI
 \end{equation}
 and investigate the sensitivity of frequency shifts on solar activity.
The slope, $b$, which measures the shift per activity index,
 and intercept, $a$,  obtained from the linear least-square fit for each
activity index, $I$, are also given in Table~3. 
The gradients obtained from the linear regression analysis show  
a higher sensitivity in the GONG data 
as compared to the MDI and is consistent with the notion that  
 the uncertainty in the determination of frequencies is inversely proportional
to the square root of the length of the time series. 
 
A  comparison of correlation coefficients for all activity indices indicates
that the best correlation is obtained for chromospheric activity indices, in
particular  $F_{10}$.  The recent analysis of low-degree modes also report 
better correlations for $F_{10}$, \ion{He}{1} equivalent width and \ion{Mg}{2} 
core-to-wing ratio for three consecutive cycles 21--23 \citep{wjc07}.
 However, these authors report relatively low correlation coefficients for 
these indices as compared to our analysis, although the length of the 
time samples  in both cases is the same, i.e. 108 days. The low-degree modes 
travel deep to the solar interior while the intermediate degree
modes are confined to the radiation and convection zones. 
Virtually all of the activity effects on the modes is at the surface. The 
surface spatial patterns of the low-$\ell$ modes are not well matched 
to the spatial distribution of surface activity, hence one may expect
intermediate- and high-degree modes to correlate better with the activity. 
We also obtain higher correlation coefficients for the 108-day GONG data sets 
compared to the  72-day MDI data. The analysis of \citet{wjc01} and \citet{sct07} 
on many different time scales show similar correlation patterns.  

Although cycle 23 has been a weak cycle compared to its predecessor,
it recorded two prominent peaks during solar maximum. The  first peak in
sunspot number (mid-2000) was higher than the secondary  peak (late-2001),
while several other solar indices (e.g., KPMI, $F_{10}$, \ion{Mg}{2}) had a 
maximum at the secondary peak.  The frequency shifts also exhibited the 
maximum  during the secondary peak. Thus $R_{\rm I}$ is the only activity proxy 
  whose maximum  does not coincide with the maximum of the frequency shifts.  

\subsubsection{ Phase-wise variation}

In order to study the correlation between frequency shifts and activity at 
different phases of the solar cycle, we divide the activity cycle into three 
phases; the rising activity period from 1996 September 22 to 1999 June 26 
({\it Phase I}),  the high-activity period from 1999 June 27 to 2003 January 12 
({\it Phase II}), and  the declining activity period from 2003 January 13 to 
2007 July 26 ({\it Phase III}). It should be noted that the relatively flat 
portion of solar activity corresponding to the beginning of minimum phase is 
not included in this analysis. The start and end dates in each activity phase are
 the same for both GONG and MDI data sets. The number of data sets representing 
 {\it Phase I, II and III}  are;  26 (GM 16 - 41),  34 (GM 44 - 77) 
 and 44 (GM 80 - 123) for GONG and 11, 17 and 22 for MDI. Our analysis 
 is restricted to {\it Phases I} and {\it II} for indices KPMI and  \ion{He}{1} 
 due to the unavailability of activity data during the entire declining phase.  
 
The calculated linear correlation coefficients, $r_P$, for both GONG and MDI
data sets for all the three phases are   shown in Figure~6. In the same figure, 
we also compare these coefficients with those obtained for the complete cycle. 
It is interesting to note that the correlation between frequency shifts and 
solar activity  changes significantly from phase to phase except for  $F_{10}$.
 In all cases, the rising and the declining phases are better correlated
 than the high-activity phase, except for CI where we obtain a decrease
 in correlation during the declining phase for the frequency shifts calculated from
GONG. A substantial decrease in correlation during high-activity phase is obtained 
for $R_{\rm I}$, MWSI, TSI and \ion{He}{1} indices. This poor correlation is also 
visible in Figures~4 and 5 where variation in $\delta\nu$ shows a different pattern 
than the activity proxies.   We point out that the $R_{\rm I}$, MWSI and TSI
 indices are most affected by the sunspots with strong magnetic fields, while 
other proxies (e.g. MPSI, \ion{Mg}{2}) are sensitive to the weak component of 
the magnetic field (see Table~2). The KPMI and  $F_{10}$ indices  are modulated by both 
components, however the dominant contribution to $F_{10}$  comes from the weak component. 
The change in correlation coefficients for \ion{He}{1} is similar to that
we obtain for CI, but the coefficients for the declining phase can not be compared 
due to insufficient measurements of the \ion{He}{1} index. It should be noted here that
these two indices are influenced by the changes occurring above the chromosphere.

Because the correlation coefficients clearly reveal a phase-wise variation, we
further investigate their relation to the corresponding linear regression slopes. 
The slopes for different phases calculated using Equation~3  are tabulated in Table~4.
It is seen that the slopes in the rising and falling phases are greater 
than that obtained for the high activity period. However, KPMI appears to
be an exceptional case where the slopes in the high activity period are marginally higher 
(within 1$\sigma$ error) than 
the rising activity period, for both GONG and MDI data sets.

 The calculated linear correlation coefficients between phase-wise 
correlation coefficients and slopes ($r_P$([$r_P(I,\delta\nu)$],[$b$])), 
and solar activity ($r_P$([$r_P(I,\delta\nu)$],[$I$])) for different activity 
indicators are summarized in Table~5 for both GONG and MDI data sets. 
 In all cases, except for CI, we obtain a significant positive correlation
between correlation coefficients and slopes, and  a negative correlation   
 between correlation coefficients and activity indices. In general, we
find that the GONG data has a higher correlation than the MDI data.  
Our results are further consistent with the findings of  \citet{sct07} where
year-wise variations in the linear-regression slopes and the correlation
coefficients are measured  from 
frequencies that were determined from time series of 9 days. 
  They also found a positive correlation between these two  but no 
 significant correlation or anti-correlation between the slopes and the solar
activity. This difference may be attributed to the fact that the frequency shifts
track the changes in solar activity much better on longer
time scales, as discussed in the previous subsection. 

\subsection{Comparison between rising and declining phases: the hysteresis
patterns}

As shown in Section~3.2.2, there is no significant difference between
correlation coefficients in rising and declining phases, except for CI, 
one may expect  that  the curves of frequency shifts plotted against
solar activity during the two separate phases should overlap with each other
or have a constant shift between them.  However, the magnetic field indices in 
cycle 22 have been found to  follow different curves during ascending and
 descending phases, exhibiting a hysteresis loop \citep{jr98, sct01}. 
It is also shown that  not all activity indices  have a linear
relationship among them  \citep{bachmann94}.
  Since cycle 23 has been referred to as {\it anomalous} solar cycle
\citep{detoma04}  due to differences in various activity measures as compared
to other cycles, we plot frequency shifts against the activity and
examine the behavior of these curves to obtain a deeper insight of the
variation of frequencies with solar activity.

In Figures~7 and 8, we show the variation of $\delta\nu$ as a function of four
different activity indices ($F_{10}$, KPMI, MPSI and  \ion {Mg}{2}) for 
GONG and MDI data sets, respectively. The open and filled symbols represent 
the ascending and descending phase, respectively;   the high activity
period is shown by {\it asterisk} symbols.  It is evident that the mode 
frequency shifts follow similar paths during the ascending and descending
phases for the radio flux, $F_{10}$, and full-disk magnetic field, KPMI. However,
the other two pairs  (MPSI, and \ion{Mg}{2}) exhibit different paths and form 
a broader curve.  The behavior of $F_{10}$, MPSI and \ion{Mg}{2} in cycle 23 
is consistent with that for cycle 22 while the variation of frequencies 
with KPMI contradicts the findings of \citet{jr98} and \citet{sct01}. We note that
the instrument at Kitt Peak was upgraded in 1992 to reduce the noise level. The
post-upgrade data were calibrated to place them at the same scale as 
pre-upgrade data but any systematic offset between the two data sets could  
lead to a constant shift in the values of KPMI and may
account for the hysteresis pattern seen in cycle 22. 

\section{Discussion}

The intermediate-degree modes are generally confined to the convection zone, 
therefore one may expect the variation in their characteristics to be 
influenced by changes in this region.  The solar cycle related shifts are believed to
occur primarily at the surface reflection of the modes. In addition, their 
variation and strong correlation with solar activity indices over the solar 
cycle provide sufficient evidence for their association with the changing 
solar magnetic activity. Various measurements have been made to estimate 
such changes in magnetic activity.  In this analysis, we have considered 
three proxies which represent quantitative measurements of the magnetic field 
strength, i.e. KPMI,  MWSI and MPSI.  As outlined earlier, the KPMI is total 
line-of-sight magnetic field  while MWSI provides only the  strong component in 
the photosphere. On the other hand, MPSI 
estimates weak fields associated with magnetic plages in the chromosphere.

In order to study the differences in the correlation between frequency shifts
and various magnetic measurements, we plot 108-day averaged photospheric fields 
against the MPSI (Figure~9). Note that the weak and strong fields do not vary 
in a similar way at all activity phases. We find a linear relationship during 
rising and falling activity periods among all three proxies, but a large deviation 
from the linear trend is seen between MPSI and MWSI during the high-activity period.
The variation in the KPMI is linear with the MPSI in rising- and high-activity
periods, however, their behavior remains unexplored during the declining phase
due to the unavailability of KPMI data. To supplement the KPMI measurements during 
the declining phase, we include in the same figure the line-of sight magnetic 
field measured by Synoptic Optical Long-term Investigations of the Sun (SOLIS)  
instrument located at Kitt Peak. The SOLIS
observations\footnote{http://solis.nso.edu/vsm/svsm\_m11\_meanfield.dat} are
made using a different spectral line (\ion{Fe}{1} 630.2 nm) than the KPMI, but
the overlapping data of these two instruments show a similar response to
active regions and other solar phenomena (J. W. Harvey, private
communication).  However, we notice a downward shift in the SOLIS fields
as compared to the KPMI which may be due to the use of a  different spectral
line.  Moreover, we find a strong correlation between SOLIS magnetic
field and the MWSI which is consistent with the results obtained for KPMI. 

Therefore,  we do not expect frequency shifts to correlate equally with all
activity indicators at different phases. Our analysis for
the complete cycle and different activity phases clearly shows that the 
frequency shifts have the best correlation  with the 10.7 cm radio flux. The 
radio flux is measured in the high chromosphere and lower corona, but it is 
modified by the evolution of sunspots (strong magnetic field) and radio plages 
(weak magnetic field). By analyzing low-$\ell$ frequencies for cycle 21--23, 
\citet{wjc07} concluded that the frequencies have good sensitivity to the 
effects of the weak component of the magnetic fields. Recently, \citet{sct07} 
have shown that activity related changes in oscillation frequencies can be 
tracked on a time scale as short as nine days. They also obtained the best 
correlation with the radio flux. In a theoretical study, \citet{camp89} also 
showed that changes in the chromospheric magnetic field strength can cause 
changes in the {\it p}-mode frequencies at higher $\ell$ ($\ge$ 100). On the 
other hand, using the technique of local helioseismology, \citet{brad00} argued 
that the magnitude of the variation in frequencies depend upon the magnetic 
field strength associated with active regions on the solar surface and the 
physical phenomenon inducing the shifts is confined to the surface layers
of the Sun. On the basis of our study, we find that both components of the 
magnetic field influence the changes in the oscillation frequencies, however 
the weak component plays a dominant role. To examine this conclusion, 
we developed various regression models between  MWSI (strong  magnetic field), 
MPSI (weak magnetic field), and $F_{10}$ and $\delta\nu$ from GONG and MDI 
data sets.  The variance for different regression models calculated using 
frequency shifts from GONG data are given Table~6. It is clearly seen that we 
obtain higher correlation  at all activity phases for the models based on either 
$F_{10}$ or combined MPSI and MWSI data. These models are able to explain 
more than 90\% of the variation in the frequencies at all  activity phases. In
contrast, the model with  strong  component of magnetic field (MWSI) explains 
 less than 50\% of the variation  while the model with only weak component  (MPSI)
explains less than 80\% of the variation in frequencies at high-activity
period. A similar behavior is also seen in case of MDI data.

\section{Summary}

The improved and continuous measurements of intermediate-degree $p$-mode 
frequencies for a complete solar cycle demonstrate that, while the frequencies
vary in phase with the solar activity, the degree of correlation between the
frequency shifts and activity indices differs from one phase of activity to
another.  

Although there is a strong correlation between frequency shifts and all
activity proxies during the rising and declining phases, we find a significant
change in the correlation at the high-activity period. The maximum
decrease is obtained for $R_{\rm I}$, MWSI, TSI, and \ion{He}{1}, where first
three indices are most affected by the sunspots with strong magnetic fields 
in the photosphere while \ion{He}{1} is influenced by the change in temperature 
in transition region. The change in correlation coefficients for CI  is 
similar to that of \ion{He}{1}  but we could not compare the coefficients
in  the declining phase  due to insufficient measurements of the \ion{He}{1} index.
 We further find a negligible change in the correlation for the 10.7 cm
radio flux which provides estimates of the variation in  both weak and strong
components of the magnetic field. It is also influenced by a slowly varying
component  ({\it S}-component)  that originates thermally in localized regions
of high electron densities and magnetic fields present in the vicinity of
sunspot and chromospheric plages. These regions remain present over
several rotations of the Sun and contribute to the change in oscillation
frequencies.  Our analysis  suggests that both components
of the magnetic field contribute significantly to the variation in  
oscillation frequencies,  however the  contribution  from the
weak component appears to be dominant. 

The correlation coefficients are strongly correlated with the slopes obtained 
from the linear regression analysis, in agreement with the earlier findings of 
\citet{sct07} using  frequencies determined
from time-series of 9 days in length. In addition, slopes and
correlation coefficients are found to be strongly anti-correlated with the
measures of solar activity.   However, the coronal index shows a poor
correlation in all cases. It may be due to the fact that the solar oscillations 
are most affected by the magnetic activity of lower atmosphere and 
are not influenced by the changes in corona. 

\acknowledgments
We thank the anonymous referee for useful comments. 
We also thank Mukul Kundu and Bill Livingston for useful discussions. 
 This work utilizes data obtained by the Global Oscillation Network
Group (GONG) project, managed by the National Solar Observatory, which
is operated by AURA, Inc. under a cooperative agreement with the
National Science Foundation. The data were acquired by instruments
operated by the Big Bear Solar Observatory, High Altitude Observatory,
Learmonth Solar Observatory, Udaipur Solar Observatory, Instituto de
Astrof\'{\i}sico de Canarias, and Cerro Tololo Interamerican
Observatory. It also utilizes data from the Solar Oscillations
Investigation/Michelson Doppler Imager on the Solar and Heliospheric 
Observatory. SOHO is a mission of international cooperation
 between ESA and NASA. NSO/Kitt Peak magnetic, and Helium measurements used
here are produced cooperatively by NSF/NOAO; NASA/GSFC and NOAA/SEL. 
SOLIS data used here are produced cooperatively by NSF/NSO and NASA/LWS. 
This study also includes data from the synoptic program at the 150-Foot 
Solar Tower of the Mt. Wilson Observatory. The Mt. Wilson 150-Foot Solar 
Tower is operated by UCLA, with funding from NASA, ONR and NSF, under 
agreement with the Mt. Wilson Institute. The unpublished solar irradiance 
data set (version v6\_001\_0804) was obtained from VIRGO Team through PMOD/WRC, 
Davos, Switzerland. This work has made use of NASA's Astronomy data 
System (ADS) and data available at NOAA's National Geophysical Data Center 
(NGDC) website. This work was supported by NASA grant NNG 05HL41I and NNG 08EI54I.


\begin{figure}
\plotone{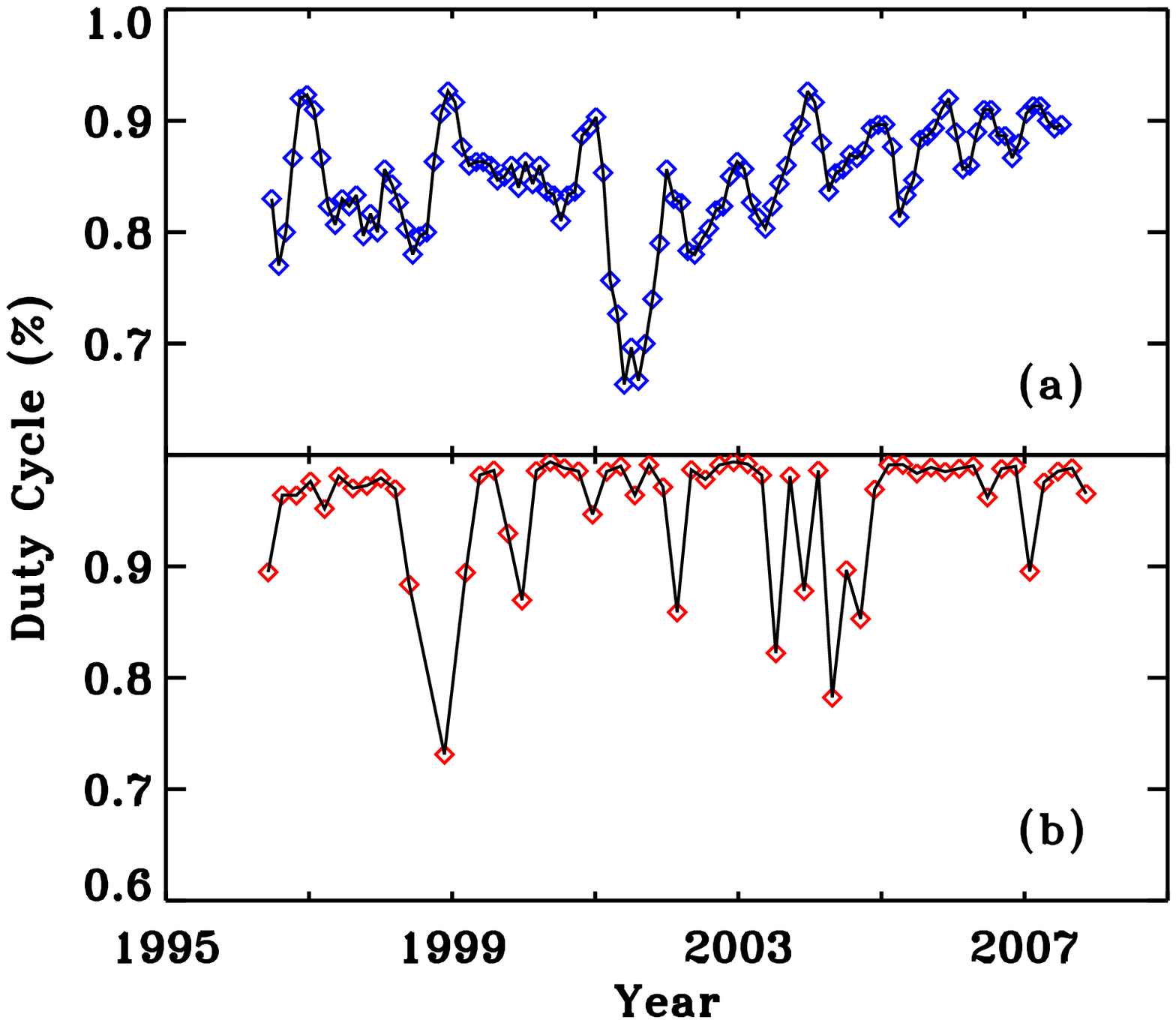}
 \caption{Duty cycles of  (a) GONG and (b) MDI time series. }
 \label{fig1}
\end{figure}
\newpage

\begin{figure}
\plotone{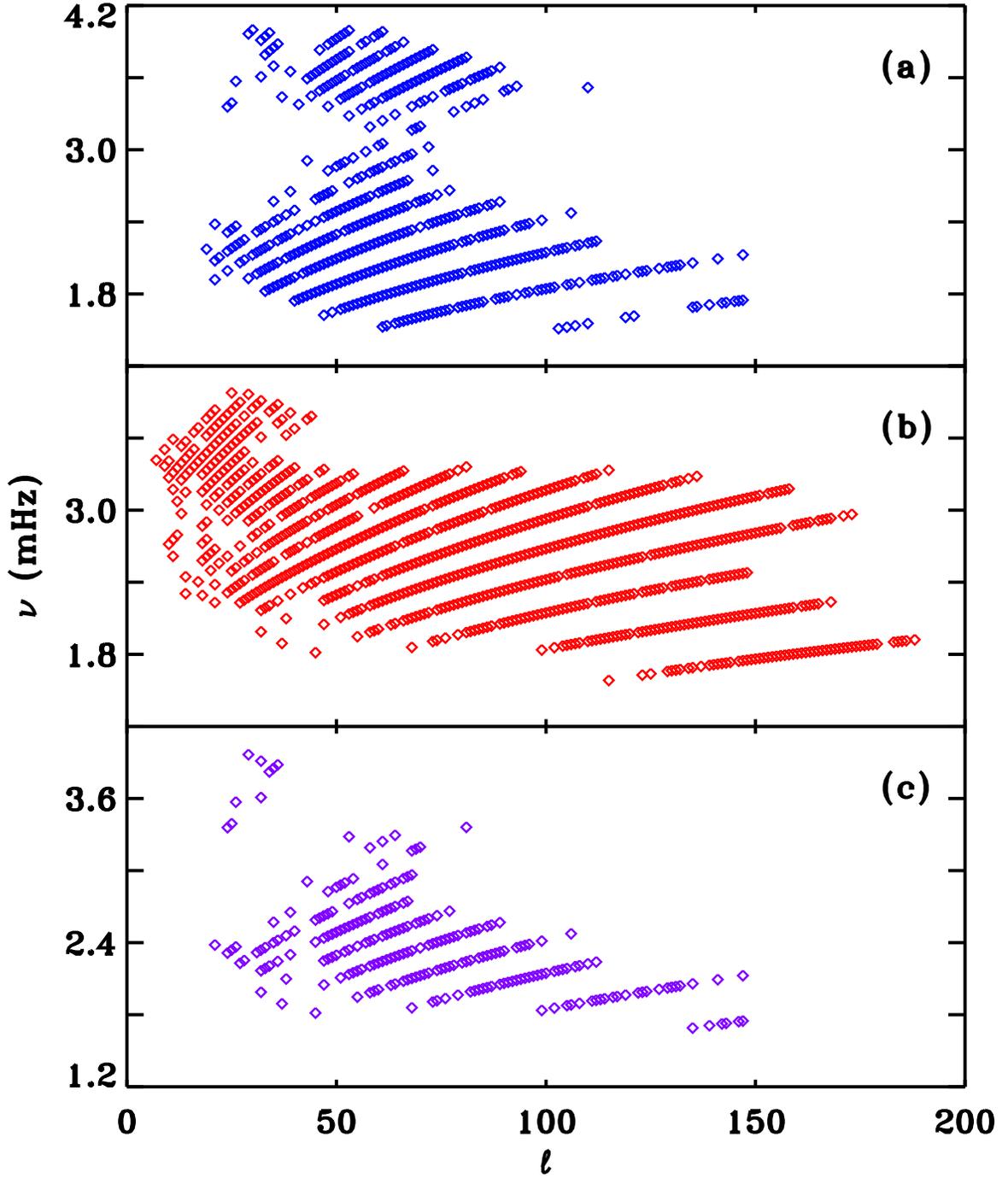}
 \caption{Comparison of common {\it p} modes observed in  all (a) GONG, (b) MDI
and (c)
  combined GONG and MDI data sets. }
 \label{fig2}
\end{figure}
\newpage

\begin{figure}
\plotone{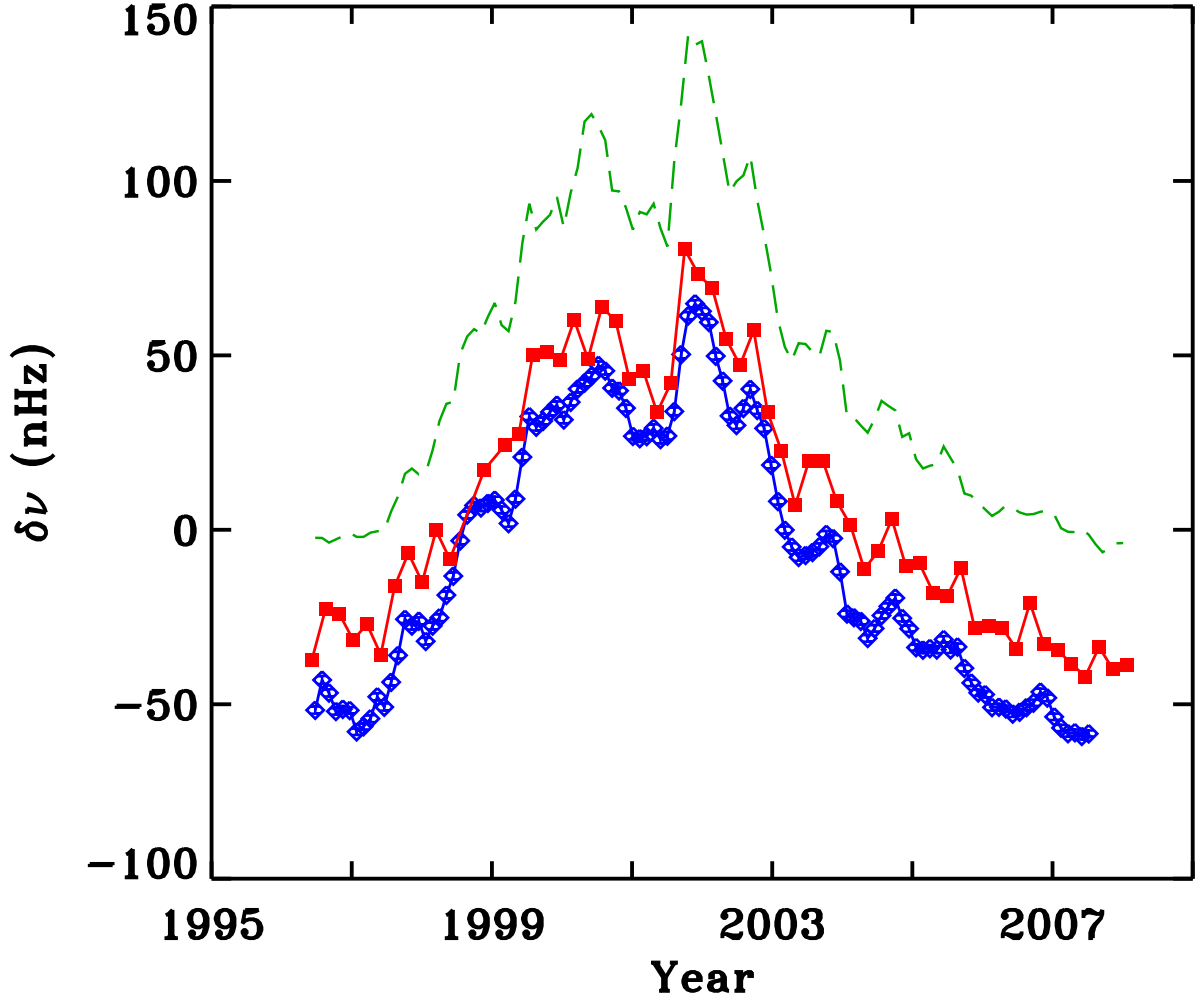}
 \caption{Temporal evolution of frequency shifts calculated from the GONG (open
symbols) and MDI (filled symbols)
  frequencies. The dashed line represents the scaled 10.7 cm radio flux.
 \label{fig3}}
\end{figure}
\newpage

\begin{figure*}
\begin{center}
\includegraphics[angle=90,scale=0.5]{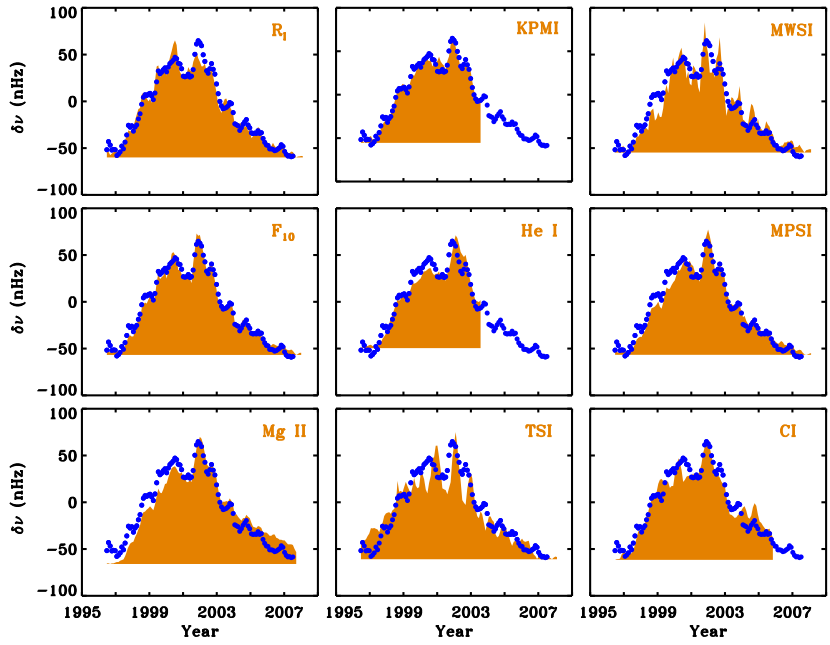}
\caption{Temporal evolution of the GONG frequency shifts (symbols) with
different activity 
proxies (filled regions).
\label{fig4}}
 \end{center}
\end{figure*}
\newpage

\begin{figure*}
\begin{center}
\includegraphics[angle=90,scale=0.5]{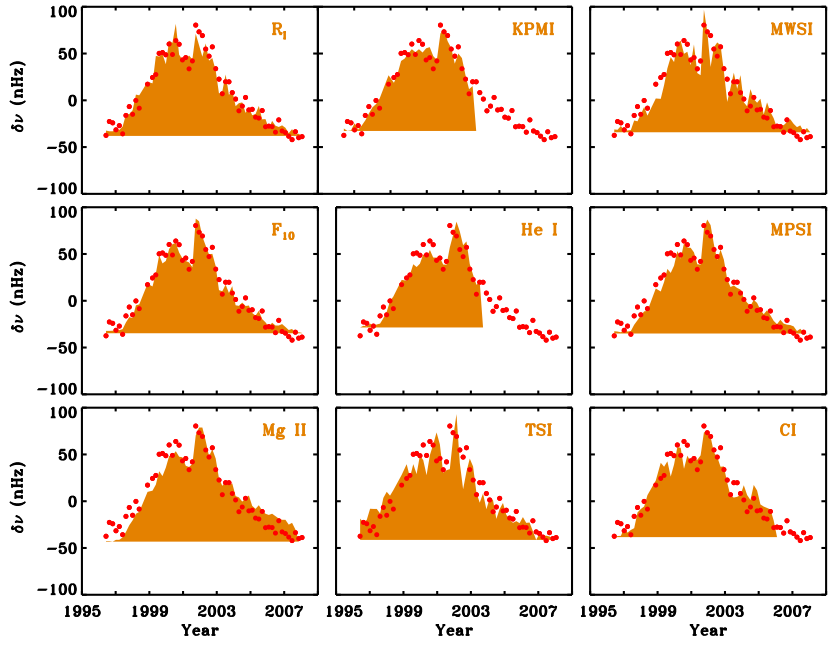}
\caption{Same as Figure~4 but for the shifts calculated from the MDI
frequencies.
\label{fig5}}
 \end{center}
\end{figure*}
\newpage
 
\begin{figure*}
\begin{center}
\includegraphics[angle=90,scale=0.5]{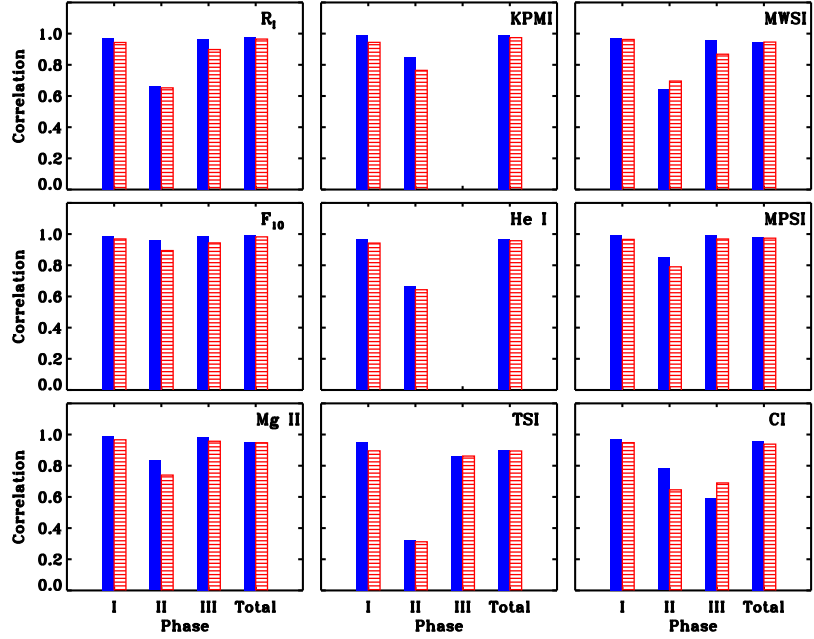}
 \caption{Bar-chart showing phase-wise variation of the Pearson's linear
correlation coefficients
 for different activity indices.  The complete solar cycle is divided into
three
 phases; the rising (I), high (II) and declining (III) activity.  Each activity
 phase has values for the GONG (left) and the MDI (right). Note that missing
values for
 KPMI and \ion{He}{1} are due to unavailability of activity measurements during
 the declining phase.\label{fig6}}
 \end{center}
\end{figure*}

\newpage

\begin{figure}
\plotone{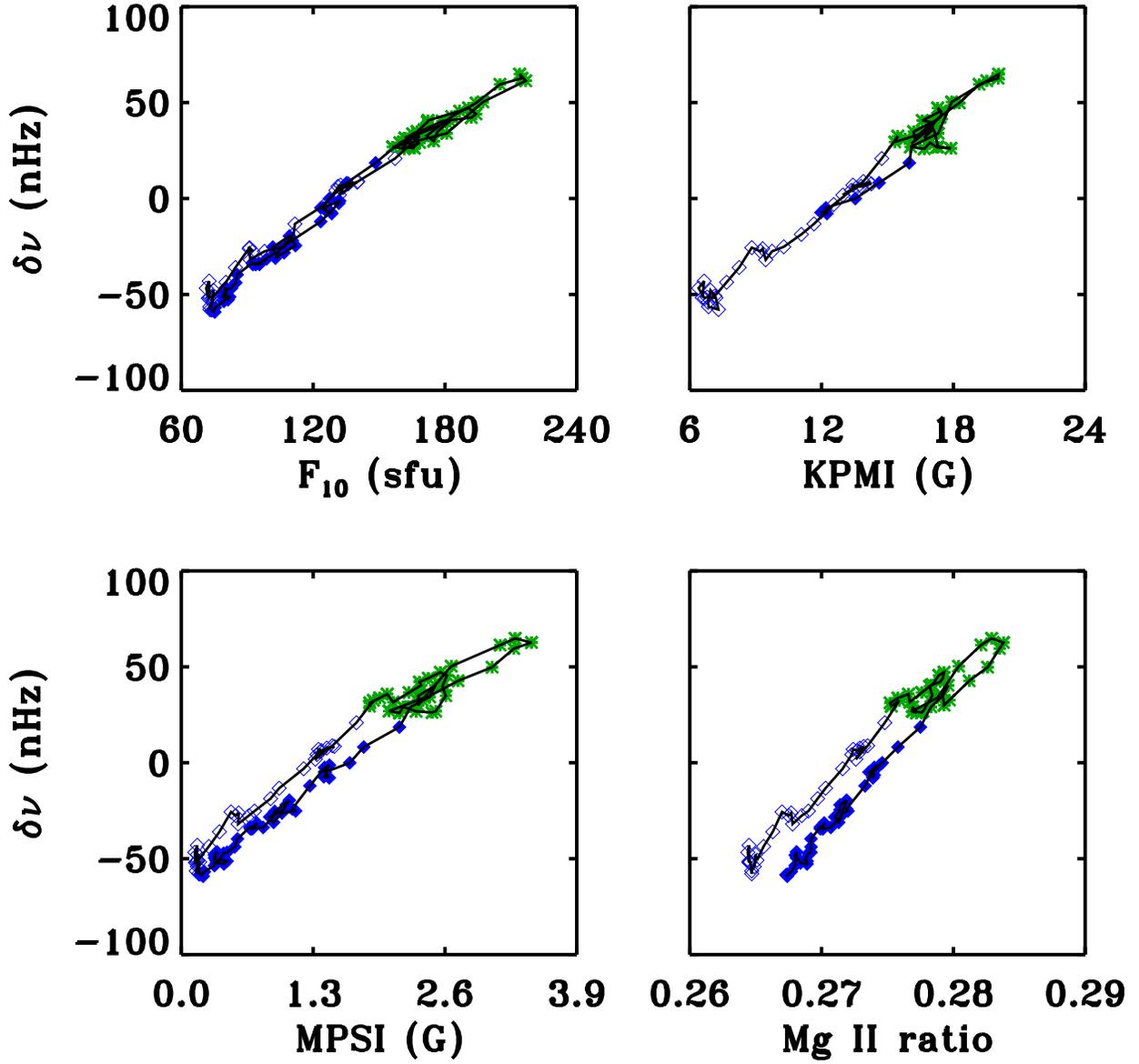}
 \caption{Variation of  GONG frequency shifts  with different activity
proxies. The rising, high and falling
 phases are shown by open symbols, asterisks and filled symbols,
respectively. The symbols have been joined by solid line to guide the
eye.   \label{fig7}}
\end{figure}
\newpage

\begin{figure}
\plotone{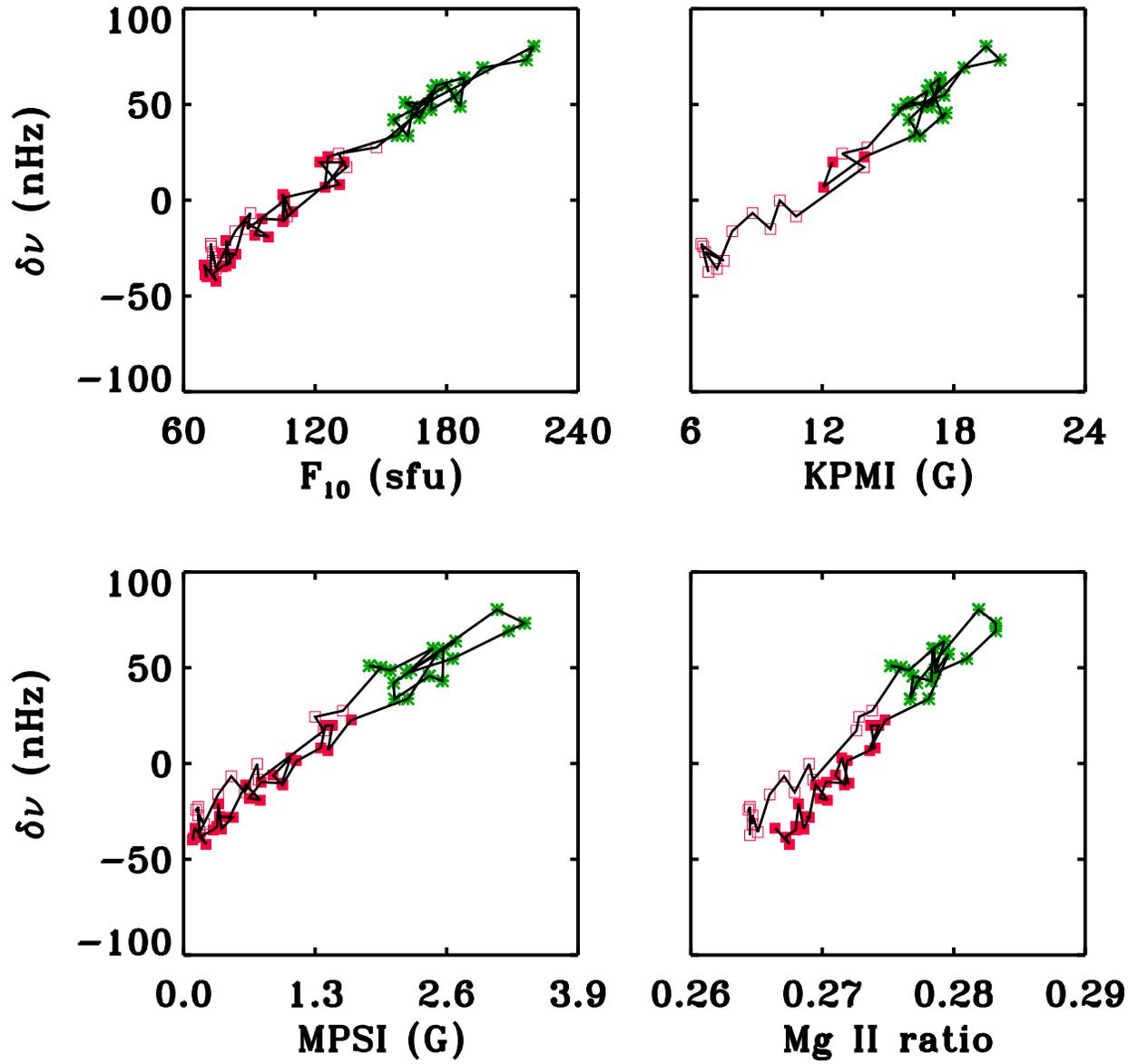}
 \caption{Same as Figure 7 but for the MDI frequency shifts.
\label{fig8}}
\end{figure}
\newpage

\begin{figure}
\plotone{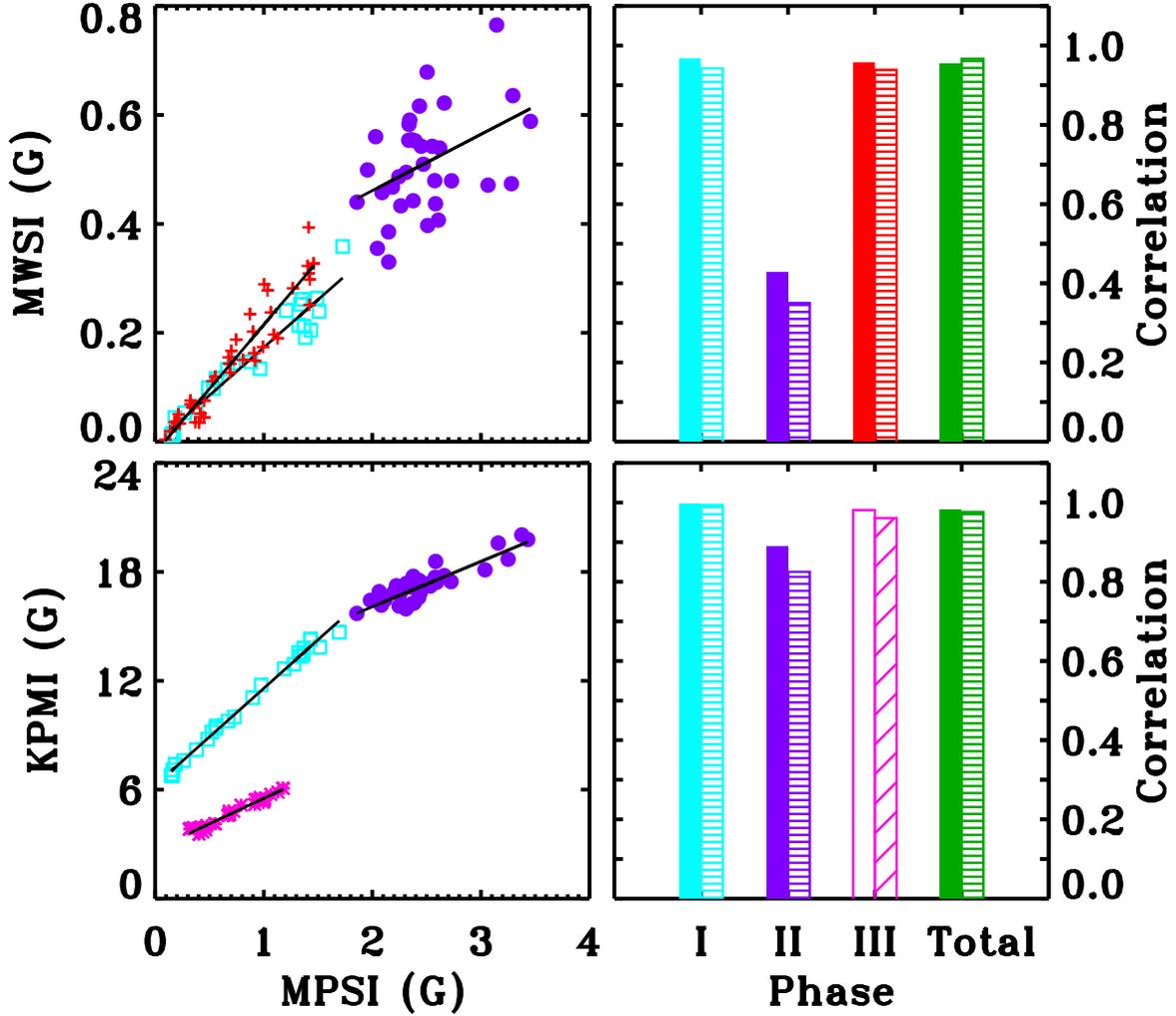}
 \caption{{\it (left)} Variation in KPMI (total field) and MWSI (strong field) as a function
of MPSI
 (weak field). Rising, high and falling activity periods are
 shown by the squares, filled circles and pluses , respectively.  The
crosses are for the 
SOLIS data. The solid lines denote the linear fits at different phases.  
{\it (right) } The linear correlation coefficients, $r_P$ (filled) and
rank correlation coefficients, $r_S$ (hatched) at different phases 
corresponding to the left panels.  In the lower panel,  $r_P$ and $r_S$
between the SOLIS measurements and MPSI are shown by open and diagonally
hatched bars, respectively. \label{fig9}}
\end{figure}


\clearpage
\begin{deluxetable}
{lll}
\tabletypesize{\scriptsize} 
\tablecaption{Description of frequency data sets.}
\tablewidth{0pt}
\tablehead{
  \colhead{}&\colhead{GONG}&\colhead{MDI}
}         
\startdata
  Period & 1996 May 1 - 2007 Aug 31& 1996 May 1 - 2008 Feb 27\\
  Number of datasets        & 113& 58\\
  Extent (days)             & 108 & 72 \\
  Frequency range (mHz)     & 1.5 $\le \nu \le $ 4.0 & 1.5 $\le \nu \le $ 4.0  
\\
  Number of common modes    & 500&  885\\
  Degree range              &19$\le \ell \le $147&  7$\le \ell \le $188 \\
  Radial nodes              & 1$\le n \le $19  &1$\le n \le $21 \\
  \enddata
\end{deluxetable}


\begin{deluxetable}
{lllll}
\tabletypesize{\scriptsize} 
\tablecaption{Various proxies for solar activity.}
\tablewidth{0pt}
\tablehead{
   \colhead{Activity} &\colhead{Spectral line}&\colhead{Influential}&
   \colhead{{End date}\tablenotemark{a}}&\colhead{Data source}\\
   \colhead{Index} &\colhead{ }&\colhead{field component} &
   \colhead{ }&\colhead{ }\\
 }         
\startdata
R$_{\rm I} $ & - &Strong &2008 27
Feb&ftp://ftp.ngdc.noaa.gov/STP/SOLAR\_DATA/SUNSPOT\_NUMBERS/ \\
KPMI  & Fe {\sc I} 868.8 nm &Both  &2003 Sep 21 &
ftp://nsokp.nso.edu/kpvt/daily/stats/mag.dat\\
MWSI      & Fe {\sc I} 525.0 nm & Strong &2008 27 Feb
&ftp://howard.astro.ucla.edu/pub/obs/mpsi\_data/index.dat\\
 F$_{10}$ & - & Both &2008 Feb 27 &
ftp://ftp.ngdc.noaa.gov/STP/SOLAR\_DATA/SOLAR\_RADIO/FLUX/\\  
He {\sc I} & He {\sc I} 1083.0 nm & -  &2003 Sep 21
&ftp://nsokp.nso.edu/kpvt/daily/stats/he.dat \\
MPSI   &Na {\sc I} 589.5 nm  & Weak & 2008 27 Feb&
ftp://howard.astro.ucla.edu/pub/obs/mpsi\_data/index.dat\\
Mg {\sc II} & Mg {\sc II} 279.9 nm  &Weak   &2007 Oct 6  &
http://www.sec.noaa.gov/ftpdir/sbuv/NOAAMgII.dat\\
 TSI   & - & - &2008 Feb 27 &
ftp://ftp.pmodwrc.ch/pub/data/irradiance/virgo/TSI\\
 CI    & Fe {\sc XIV} 530.3 nm   & - &2005 Dec 31 &
ftp://ftp.ngdc.noaa.gov/STP/SOLAR\_DATA/SOLAR\_CORONA/\\
 \enddata
 \tablenotetext{a}{Start date for all data sets is 1996 May 1.  }
\end{deluxetable}


\begin{deluxetable}
{lllllrr}
\tabletypesize{\scriptsize} 
\tablecaption{The correlation coefficients and the linear regression parameters 
between activity indices ($I$) and frequency shift ($\delta\nu$) for the total
period covered in this analysis.
Shown here are the Pearson's linear coefficient ($r_P$), Spearman's rank
correlation ($r_S$),
the intercept ($a$) and the mean shift per unit change in activity ($b$).
The two-sided significance ($P_S$) is smaller than 10$^{-16}$ and is not shown
here.}
\tablewidth{0pt}
\tablehead{
  \colhead{Activity Index}& \colhead{}& \colhead{Data Source}&
\colhead{$r_P$($I$,$\delta\nu$)}& \colhead{$r_S$($I$,$\delta\nu$)} &
\colhead{$a$}& \colhead{$b$}\\
   \colhead{ }& \colhead{}& \colhead{ }& \colhead{ }& \colhead{ } &
\colhead{($\mu$Hz)}& \colhead{($\mu$Hz per activity unit)}\\
}         
\startdata
 $R_I$   &&GONG&0.98&0.98&$-$6.28$\times$10$^{-2}$ $\pm$
1.60$\times$10$^{-4}$&9.07$\times$10$^{-4}$ $\pm$ 2.28$\times$10$^{-6}$ \\
         &&MDI&0.97&0.95&$-$4.01$\times$10$^{-2}$ $\pm$
2.25$\times$10$^{-4}$&8.31$\times$10$^{-4}$ $\pm$ 3.22$\times$10$^{-6}$\\
 \\
KPMI && GONG&0.99&0.96&$-$1.11$\times$10$^{-2}$ $\pm$
4.09$\times$10$^{-4}$&8.63$\times$10$^{-3}$ $\pm$ 2.87$\times$10$^{-5}$ \\
     && MDI&0.98&0.93&$-$8.47$\times$10$^{-2}$ $\pm$ 5.92$\times$10$^{-4}$&
8.02$\times$10$^{-3}$ $\pm$ 4.16$\times$10$^{-5}$ \\
 \\
MWSI &&GONG&0.95&0.95&$-$5.54$\times$10$^{-2}$ $\pm$
1.48$\times$10$^{-4}$&1.83$\times$10$^{-1}$ $\pm$ 4.75$\times$10$^{-4}$ \\
     &&MDI&0.95& 0.94&$-$3.43$\times$10$^{-2}$ $\pm$
2.10$\times$10$^{-4}$&1.67$\times$10$^{-1}$ $\pm$ 6.59$\times$10$^{-4}$\\
 \\
$F_{10}$  &&GONG&0.99&0.98&$-$1.17$\times$10$^{-1}$ $\pm$
2.80$\times$10$^{-4}$&8.76$\times$10$^{-4}$ $\pm$ 2.18$\times$10$^{-6}$  \\
          &&MDI&0.98&0.97&$-$9.11$\times$10$^{-2}$ $\pm$
3.95$\times$10$^{-4}$&8.14$\times$10$^{-4}$ $\pm$ 3.10$\times$10$^{-6}$\\
 \\
He {\sc I}  &&GONG&0.97&0.95&$-$1.84$\times$10$^{-1}$ $\pm$
6.59$\times$10$^{-4}$&2.89$\times$10$^{-3}$ $\pm$ 9.81 $\times$10$^{-6}$  \\
             &&MDI&0.96& 0.95&$-$1.55$\times$10$^{-1}$ $\pm$
9.58$\times$10$^{-4}$&2.71$\times$10$^{-3}$ $\pm$ 1.42 $\times$10$^{-5}$\\
 \\
MPSI &&GONG&0.98&0.97&$-$5.98$\times$10$^{-2}$ $\pm$
1.54$\times$10$^{-4}$&3.97$\times$10$^{-2}$ $\pm$ 1.00$\times$10$^{-4}$  \\
     &&MDI&0.98&0.97&$-$3.83$\times$10$^{-2}$ $\pm$
2.19$\times$10$^{-4}$&3.71$\times$10$^{-2}$ $\pm$ 1.43$\times$10$^{-4}$ \\
 \\
Mg {\sc II} &&GONG&0.95& 0.94&$-$1.92 $\pm$ 4.96$\times$10$^{-3}$&7.01 $\pm$
1.82 $\times$10$^{-2}$ \\
             &&MDI&0.95&0.93&$-$1.75 $\pm$ 7.22$\times$10$^{-3}$&6.49 $\pm$
2.65 $\times$10$^{-2}$ \\
 \\
 TSI &&GONG&0.91&0.89&$-$108.83 $\pm$
2.95$\times$10$^{-1}$&7.97$\times$10$^{-2}$ $\pm$ 2.16$\times$10$^{-4}$  \\
     &&MDI&0.91& 0.90&$-$104.31 $\pm$
4.32$\times$10$^{-1}$&7.64$\times$10$^{-2}$ $\pm$ 3.16$\times$10$^{-4}$\\
  \\
CI  &&GONG&0.96& 0.94&$-$8.37$\times$10$^{-2}$ $\pm$
2.69$\times$10$^{-4}$&1.18$\times$10$^{-2}$ $\pm$ 3.59$\times$10$^{-5}$ \\
    &&MDI&0.94& 0.92&$-$5.84$\times$10$^{-2}$ $\pm$
3.88$\times$10$^{-4}$&1.08$\times$10$^{-2}$ $\pm$ 5.17$\times$10$^{-5}$\\
\enddata
\end{deluxetable}


\begin{deluxetable}
{llrrrrrrrrrrrrrrrrrr}
\tabletypesize{\scriptsize} 
\tablecaption{Calculated linear regression slopes (in $\mu$Hz per activity
unit) for different activity phases of solar cycle. 
Note that the slopes for the total period are different from Table 3 as these
are calculated
for a different period. Here periods covered for GONG and MDI are same in all
activity periods.
Missing values of KPMI and \ion{He}{1} for {\it Phase III} are due to the
unavailability of activity 
measurements.}
\tablewidth{0pt}
\tablehead{
  \colhead{Activity}& \colhead{Data}& \colhead{Phase I}& \colhead{Phase II}&
\colhead{Phase III}& \colhead{Total}\\
  \colhead{Index}& \colhead{Source}& \colhead{ }& \colhead{ }& \colhead{ }&
\colhead{ }\\
}
\startdata
 $R_I$   &GONG & 8.56$\times$10$^{-4}$ $\pm$ 4.27$\times$10$^{-5}$&
5.72$\times$10$^{-4}$ $\pm$ 1.12$\times$10$^{-4}$&
  9.26$\times$10$^{-4}$ $\pm$ 3.74$\times$10$^{-5}$&9.07$\times$10$^{-4}$ $\pm$
1.74$\times$10$^{-5}$\\
    &MDI & 6.66$\times$10$^{-4}$ $\pm$ 7.39$\times$10$^{-5}$&
5.56$\times$10$^{-4}$ $\pm$ 1.62$\times$10$^{-4}$&
  8.82$\times$10$^{-4}$ $\pm$ 9.43$\times$10$^{-5}$&8.35$\times$10$^{-4}$ $\pm$
3.17$\times$10$^{-5}$\\
 \\
 KPMI &GONG & 8.56$\times$10$^{-3}$ $\pm$ 2.72$\times$10$^{-4}$&
8.69$\times$10$^{-3}$ $\pm$ 9.49$\times$10$^{-4}$&
  - &8.77$\times$10$^{-3}$ $\pm$ 1.66$\times$10$^{-4}$\\
 &MDI & 7.40$\times$10$^{-3}$ $\pm$ 8.19$\times$10$^{-4}$&
7.96$\times$10$^{-3}$ $\pm$ 1.68$\times$10$^{-4}$&
  - &8.06$\times$10$^{-3}$ $\pm$ 3.35$\times$10$^{-4}$\\
 \\
 MWSI&GONG & 2.63$\times$10$^{-1}$ $\pm$ 1.33$\times$10$^{-2}$&
7.65$\times$10$^{-2}$ $\pm$ 1.57$\times$10$^{-2}$&
  1.68$\times$10$^{-1}$ $\pm$ 7.99$\times$10$^{-3}$& 1.79$\times$10$^{-1}$
$\pm$ 5.86$\times$10$^{-3}$\\
 &MDI &2.19$\times$10$^{-1}$ $\pm$ 1.95$\times$10$^{-2}$& 8.28$\times$10$^{-2}$
$\pm$ 2.14$\times$10$^{-2}$&
  1.59$\times$10$^{-1}$ $\pm$ 1.98$\times$10$^{-2}$& 1.66$\times$10$^{-1}$
$\pm$ 7.97$\times$10$^{-3}$\\ 
\\
 $F_{10}$  & GONG &9.34$\times$10$^{-4}$ $\pm$
3.10$\times$10$^{-5}$&6.27$\times$10$^{-4}$ $\pm$ 3.24$\times$10$^{-5}$&
 9.54$\times$10$^{-4}$ $\pm$ 2.34$\times$10$^{-5}$ &8.71$\times$10$^{-4}$ $\pm$
1.12$\times$10$^{-5}$  \\
 & MDI  &7.82$\times$10$^{-4}$ $\pm$
6.32$\times$10$^{-5}$&6.16$\times$10$^{-4}$ $\pm$ 7.71$\times$10$^{-5}$&
 9.51$\times$10$^{-4}$ $\pm$ 7.29$\times$10$^{-5}$ &8.10$\times$10$^{-4}$ $\pm$
2.16$\times$10$^{-5}$  \\
 \\
 He {\sc I} & GONG  &2.99$\times$10$^{-3}$ $\pm$
1.57$\times$10$^{-4}$&1.52$\times$10$^{-3}$ $\pm$ 2.96$\times$10$^{-4}$&
 -&2.88$\times$10$^{-3}$ $\pm$ 9.80$\times$10$^{-5}$\\
 &MDI   &2.57$\times$10$^{-3}$ $\pm$
2.89$\times$10$^{-4}$&1.58$\times$10$^{-3}$ $\pm$ 4.71$\times$10$^{-4}$&
 -&2.69$\times$10$^{-3}$ $\pm$ 1.45$\times$10$^{-4}$\\
\\
 MPSI&GONG & 4.61$\times$10$^{-2}$ $\pm$
1.42$\times$10$^{-3}$&2.44$\times$10$^{-2}$ $\pm$ 2.65$\times$10$^{-3}$&
 4.16$\times$10$^{-2}$ $\pm$ 9.54$\times$10$^{-4}$&3.96$\times$10$^{-2}$ $\pm$
7.92$\times$10$^{-4}$\\
 &MDI& 4.02$\times$10$^{-2}$ $\pm$ 3.40$\times$10$^{-3}$&2.31$\times$10$^{-2}$
$\pm$ 4.48$\times$10$^{-3}$&
 4.15$\times$10$^{-2}$ $\pm$ 2.32$\times$10$^{-3}$&3.69$\times$10$^{-2}$ $\pm$
1.21$\times$10$^{-3}$\\
 \\
 Mg {\sc II} &GONG &7.14 $\pm$2.15$\times$10$^{-1}$&4.31
$\pm$4.98$\times$10$^{-1}$
 &8.03 $\pm$2.06$\times$10$^{-1}$&7.10 $\pm$2.15$\times$10$^{-1}$\\
 &MDI&6.08 $\pm$5.13$\times$10$^{-1}$&4.06 $\pm$9.23$\times$10$^{-1}$
 &7.93 $\pm$5.25$\times$10$^{-1}$&6.60 $\pm$3.14$\times$10$^{-1}$\\
\\
TSI&GONG&9.47$\times$10$^{-2}$ $\pm$
6.12$\times$10$^{-3}$&1.50$\times$10$^{-2}$ $\pm$ 7.60$\times$10$^{-3}$&
7.48$\times$10$^{-2}$ $\pm$ 6.81$\times$10$^{-3}$&8.03$\times$10$^{-2}$ $\pm$
3.74$\times$10$^{-3}$\\
&MDI&8.03$\times$10$^{-2}$ $\pm$ 1.26$\times$10$^{-2}$&1.56$\times$10$^{-2}$
$\pm$ 1.18$\times$10$^{-2}$&
6.94$\times$10$^{-2}$ $\pm$ 8.91$\times$10$^{-3}$&7.52$\times$10$^{-2}$ $\pm$
5.26$\times$10$^{-3}$\\
 \\
 CI&GONG&9.46$\times$10$^{-3}$ $\pm$
4.78$\times$10$^{-4}$&7.87$\times$10$^{-3}$ $\pm$ 1.10$\times$10$^{-4}$&
 9.33$\times$10$^{-3}$ $\pm$ 2.47$\times$10$^{-3}$&1.21$\times$10$^{-2}$ $\pm$
3.92$\times$10$^{-4}$\\
 &MDI&7.95$\times$10$^{-3}$ $\pm$ 8.44$\times$10$^{-4}$&6.65$\times$10$^{-3}$
$\pm$ 1.97$\times$10$^{-4}$&
 1.04$\times$10$^{-2}$ $\pm$ 3.01$\times$10$^{-3}$&1.11$\times$10$^{-2}$ $\pm$
6.18$\times$10$^{-4}$\\
 
 \\
 \enddata
\end{deluxetable}

\begin{deluxetable}
{lllllll}
\tabletypesize{\scriptsize} 
\tablecaption{The correlation statistics between phase-wise correlation
coefficients and slopes
($r_P$([$r_P(I,\delta\nu)$],[$b$])), and solar activity
($r_P$([$r_P(I,\delta\nu)$],[$I$])) for different activity indicators using
GONG and MDI data sets.}
\tablewidth{0pt}
\tablehead{
\colhead{}&\colhead{}&\multicolumn{2}{c}{$r_P$([$r_P(I,\delta\nu)$],[$b$])}&&
\multicolumn{2}{c}{$r_P$([$r_P(I,\delta\nu)$],[$I$])}\\
\cline{3-4} \cline{6-7}\\
  \colhead{Activity Index}& \colhead{}& \colhead{GONG}& \colhead{MDI}&&
\colhead{GONG}& \colhead{MDI}\\
}
\startdata
 $R_I$   && 0.98 &0.74 &&$-$0.92&$-$0.85\\
  KPMI  &&0.90 &0.74 &&$ -$0.85 &$-$0.77\\
 MWSI &&0.85 &0.94&& $-$0.95 &$-$0.85\\
 $F_{10}$ &&0.93 &0.87& &$-$0.89 &$-$0.72 \\
 He {\sc I}  &&0.99 &0.99&&$ -$0.85 &$-$0.82 \\
  MPSI &&0.97 &0.97&& $-$0.96 &$-$0.92 \\
 Mg {\sc II} &&0.87& 0.97& & $-$0.98 &$-$0.95 \\
 TSI&& 0.99 &0.99 && $-$0.85 &$-$0.87\\
 CI&&0.47 &0.34&& $-$0.05 &$-$0.53\\
  \enddata
\end{deluxetable}

\begin{deluxetable}
{lrrrrrrrrrrrrrrrrrr}
\tabletypesize{\scriptsize} 
\tablecaption{Calculated variance for different regression models using GONG
frequency shifts ($\delta\nu$) and activity indices ($F_{10}$, MPSI and MWSI)
for different phases of the solar cycle. The coefficients $A$, $B$ and $C$ are
determined from the multiple linear regression fit. }
\tablewidth{0pt}
\tablehead{
 \colhead{Regression Model}& \colhead{Total}& \colhead{Phase I}& \colhead{Phase
II}& \colhead{Phase III}\\
}
\startdata
 $\delta\nu=A+B$$ F_{10}$ &    0.99 &0.98  &   0.96  &   0.99  \\

$\delta\nu=A+B$ (MPSI+MWSI)&  0.97 & 0.99 & 0.92 & 0.99 \\

 $\delta\nu=A+B$ MPSI+ $C$ MWSI&  0.97& 0.99 & 0.93 & 0.99\\

 $\delta\nu=A+B$ MPSI&    0.96 & 0.99 &   0.77 &   0.98  \\

$\delta\nu=A+B$ MWSI&   0.90 &  0.93 &    0.47  &   0.84 \\
 \enddata
\end{deluxetable}



\begin{thebibliography}{}

\bibitem[Antia(2003)]{antia03}
Antia, H. M. 2003, \apj, 590, 567

\bibitem[Bachmann \& Brown(1993)]{bachmann93} 
 Bachmann, K. T., \&  Brown, T. M. 1993, \apj,  411, L45
 
\bibitem[Bachmann \& White(1994)]{bachmann94} 
 Bachmann, K. T., \&  White, O. R. 1994, \solphys, 150, 347

 \bibitem[Campbell \& Roberts (1989)]{camp89} Campbell, W. R., \& 
Roberts, B. 1989, \apj, 338, 538

\bibitem[Chaplin et al.(2001)]{wjc01} Chaplin, W. J., Elsworth, Y.,
 Isaak, G. R., Marchenkov, K. I., Miller, B. A., \&  New, R. 2001,
 \mnras, 322, 22
 
\bibitem[Chaplin et al.(2007)]{wjc07} Chaplin, W. J., Elsworth, Y., Miller, 
B. A., \&  Verner, G. A. 2007, \apj, 659, 1760
 
\bibitem[Christensen-Dalsgaard \& Berthomieu (1991)]{jcd91}
Christensen-Dalsgaard,  J., \& Berthomieu, J. 1991, in Solar Interior and
Atmosphere,  eds. A. N. Cox, W. C. Livingston, M. Matthews 
(Tucson: University of Arizona Press), 401

\bibitem[Covington(1969)]{covington69} 
Covington, A. E. 1969, \jrasc, 63, 125

\bibitem[de Toma et al.(2004)]{detoma04} de Toma, G., White, O. R., Chapman, G.
A., Walton, S. R., Preminger, D. G., \&  Cookson, A. M. 2004, \apj, 609, 1140

\bibitem[Dziembowski \& Goode(2005)]{dzi05} Dziembowski, 
W. A.,  \& Goode, P. R. 2005, \apj, 625, 548

\bibitem[Fr{\"o}hlich et al.(1997)]{cf97} 
Fr{\"o}hlich, C., Crommelynck, D., Wehrli, C., Anklin, M., Dewitte, S., Fichot,
 A., Finsterle, W., Jim{\'e}nez, A., Chevalier, A., \&  Roth, H. J. 1997,
 \solphys, 175, 267
 
\bibitem[Harvey(1984)]{harvey84} Harvey, J.~W. 1984, NASA, 
Washington  Solar Irradiance Variations on Active Region Time Scales 
(SEE N84-27635 17-92), 197 

\bibitem[Hindman et al.(2000)]{brad00} Hindman, B., Haber, D., Toomre, J.,  \& 
Bogart, R. S. 2000, \solphys, 192, 363

\bibitem[Howe et al.(1999)]{howe99} Howe, R., Komm, R.,  \&  Hill, 
F. 1999, \apj, 524, 1084

\bibitem[Howe et al.(2002)]{howe02} Howe, R., Komm, R.,  \&  Hill, 
F. 2002, \apj, 580, 1172

\bibitem[Jain \& Bhatnagar(2003)]{jain03} Jain, K.,  \& 
Bhatnagar, A. 2003, \solphys, 213, 257

\bibitem[Jain et al.(2000)]{jain00} Jain, K., Tripathy, S. C.,  \& 
Bhatnagar, A. 2000, \apj, 542, 521

\bibitem[Jim\'enez-Reyes et al.(1998)]{jr98} Jim\'enez-Reyes, S. J., R\'egule,
C., Palle, P. L.,  \&  Roca Cort\'es, T. 1998, \aap, 329, 1119 

\bibitem[Kundu (1965)]{kundu65} Kundu, M. R. 1965, {\it Solar Radio Astronomy}
(New York: Interscience Publication)

\bibitem[Livingston et al.(1976)]{livingston76} Livingston, W.~C., 
Harvey, J., Slaughter, C., \&  Trumbo, D. 1976, \ao, 15, 40 

\bibitem[Rabello-Soares, Korzennik, \& Schou(2006)]{rs06}
 Rabello-Soares, M. C.,  Korzennik, S. G., \&   Schou, J. 2006,
in ESA SP-624, Proceedings of SOHO18/GONG 2006/HELAS I, Beyond the Spherical
Sun, eds. K. Fletcher \& M. J. Thompson (Noordwijk: ESA), 71

\bibitem[Rybansky et al.(1994)]{Rybansky94} Rybansky, M.,
 Rusin, V., Minarovjech, M., \&  Gaspar, P. 1994, \solphys, 152, 153

\bibitem[Salabert et al.(2004)]{david04}
Salabert, D., Fossat, E., Gelly, B., Kholikov, S., Grec, G., Lazrek, M.,
  \& Schmider, F. X. 2004,  \aap, 413, 1135
 
\bibitem[Schou (1992)]{schou92}  Schou, J. 1992, Ph.D. Thesis, Aarhus
University, Aarhus, Denmark. 
 
\bibitem[Tripathy et al.(2001)]{sct01}  Tripathy, S. C., Kumar, B.,
 Jain, K., \&  Bhatnagar, A. 2001, \solphys, 200, 3
 
\bibitem[Tripathy et al.(2007)]{sct07}  Tripathy, S. C., Hill, F., 
Jain, K.,  \& Leibacher, J. W. 2007, \solphys, 243, 105

 \bibitem[Ulrich (1991)]{ulrich91} Ulrich, R. K. 1991, Adv. Space Res., 11, 217

\bibitem[Viereck et al.(2001)]{Viereck01} Viereck, R.,
 Puga, L., McMullin, D., Judge, D., Weber, M., \&  Tobiska, W. K. 2001, \grl,
28, 1343
 
 \bibitem[Woodard \& Noyes (1985)]{woo85} Woodard, M. F., \& 
 Noyes, R. W. 1985, Nature, 318, 449
  
  \bibitem[Woodard et al. (1991)]{woo91} Woodard, M. F., Libbretch, K. G.,
  Kuhn, J., \& Murray, N. 1991, \apj, 373, L81

\end{thebibliography}
\end{document}